# Diffraction of collinear correlated photon pairs by an ultrasonic wave

**Piotr Kwiek**

*University of Gdansk, Institute of Experimental Physics, Wita Stwosza 57, 80-952 Gdańsk, Poland*
*e-mail address: fizpk@ug.edu.pl*

The phenomenon of collinear correlated photon pairs diffraction by an ultrasonic wave is investigated for Bragg incidence. A BBO crystal was used for producing collinear correlated photon pairs via type-I spontaneous parametric down-conversion (SPDC I-type). It is shown experimentally that the Bragg angle for photon pairs diffraction is identical to the one corresponding to single photons diffraction. The numbers of single photons and photon pairs counts in discrete diffraction orders were measured as functions of the Raman-Nath parameter. Similarly, the number of coincidence photon counts in separate diffraction orders was also investigated. What is more, simple analytical formulas are derived which perfectly describe experimentally obtained data.

OCIS codes: (050.1940) Diffraction; (230.1040) Acousto-optical devices; (230.4110) Modulators; (270.4180) Multiphoton processes.

1. **Introduction**

Acousto-optic devices are commonly used to modify properties of optical beams such as amplitude or intensity, phase, frequency, polarization and direction of propagation [1, 2]. Depending on which property of light is manipulated, one can name intensity modulators, phase modulators, frequency modulators, polarization modulators and spatial light modulators. Acousto-optic devices were also employed in experimental investigations devoted to photon pairs or entangled photon pairs [3-8]. Here, in most cases, acousto-optic modulators altered, at a time, parameters of just one of the photons of a given pair. Only the works of the group of Zeilinger [7] and Kwiek [8] considered simultaneous photon pairs interaction with an ultrasonic wave. The investigation of Zeilinger group concerned interaction of non-collinear entangled photon pairs with ultrasonic wave. While the research performed by Kwiek was obtained for collinear photon pairs. These included Raman-Nath and intermediate light diffraction region [8]. In the experiment carried out by Zeilinger group the pair of non-collinear photons from the SPDC I-type source were incident onto the ultrasonic wave, one of them at the positive and the other at the negative Bragg angles. In the present paper we discuss, not investigated so far, interaction phenomenon of collinear correlated photon pairs with an ultrasonic wave, within the Bragg diffraction regime. We precede the description of photons interaction with an ultrasonic wave with a short introduction which covers the principles of light diffraction by ultrasound.

2. **Light Diffraction by Ultrasonic Wave**

The phenomenon of light diffraction by an ultrasonic wave was discovered experimentally in 1932 by P. Debye and F. W. Sears [9] in the USA and, independently, by R. Lucas and P. Biquard [10] in France. The best-known theory of light diffraction by an ultrasonic wave was developed by C. V. Raman and N. S. Nagendra Nath within 1935-1936 and was published in a series of five common papers [11], followed by one individual by Nath [12].

Nowadays, two parameters are commonly used to describe diffraction of light by an ultrasonic wave [13,14]:

Raman-Nath parameter $\quad v = \dfrac{2\pi \mu_1 L}{\lambda}$, $\qquad$ (1)

and Klein-Cook parameter $\quad Q = \dfrac{2\pi \lambda\, L}{\mu_0 \Lambda^2}$, $\qquad$ (2)

where $\lambda$ is the wavelength of the (incident) light wave in a vacuum, $\Lambda$ denotes the wavelength of the ultrasonic wave in the considered medium, $\mu_0$ stands for the refractive index of the (undisturbed) medium in which the ultrasonic wave propagates, and $\mu_1$ is the maximum variation of the refractive index which is proportional to the sound pressure amplitude.

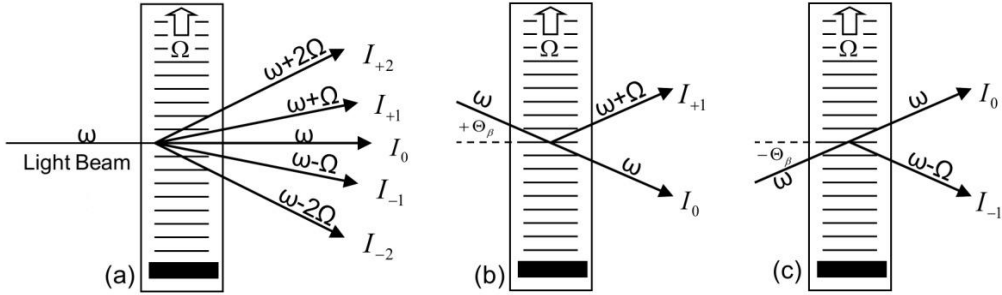

FIG. 1. Schematic representation of light diffraction by an ultrasonic wave. The frequency of light in separate diffraction orders, resulting from the Doppler effect, depends on the number of a diffraction order and the mutual propagation directions of light and ultrasonic wave. The figure illustrates three different situations:
a) Raman-Nath diffraction in case of normal incidence of light on an ultrasonic wave,
b) Bragg diffraction in case of $\Theta_B > 0$,
c) Bragg diffraction for $\Theta_B < 0$.

The Klein-Cook parameter divides light diffraction by an ultrasonic wave into three regimes, in which the observed phenomena differ fundamentally [13]. $Q \ll 1$ defines the so-called Raman and Nath region and diffraction phenomenon within this zone is called Raman-Nath diffraction. Occurrence of numerous diffraction orders, both for normal and oblique light incidence on ultrasonic wave, is a distinctive feature of Raman-Nath diffraction. The intensity $I_n$ of diffracted light in the n-th diffraction order for normal incidence of light on an ultrasonic wave can be described in this region by a corresponding Bessel function of the first kind:

$$I_n = J_n^2(v), \quad n = 0, \pm 1, \pm 2, \pm 3 \ldots,  \qquad (3)$$

where $v$ is the previously defined Raman-Nath parameter (1).

In the Bragg diffraction regime, when $Q \gg 1$, one can observe only two diffraction orders (Fig. 1): the zero-th and plus first one or the zero-th and minus first one, and strong acousto-optic interaction takes place when light incidents ultrasonic wave at an angle $\Theta_B$ satisfying the Bragg condition:

$$\sin\Theta_B = \pm \frac{\lambda}{2\mu_0 \Lambda}. \qquad (4)$$

The following formulae, describing diffracted light intensities in the zero-th $I_0$ and plus first $I_1$ diffraction order (when $\Theta_B > 0$) or in the zero-th and minus first $I_{-1}$ diffraction order (when $\Theta_B < 0$), were derived in 1956 by P. Phariseau [15] from the Raman-Nath equation set [12]:

$$I_0 = \phi_0 \cdot \phi_0^* = \cos^2(\frac{v}{2}), \qquad I_1 = \phi_1 \cdot \phi_1^* = \sin^2(\frac{v}{2}), \qquad I_{-1} = \phi_{-1} \cdot \phi_{-1}^* = \sin^2(\frac{v}{2}), \qquad (5)$$

where

$$\phi_0 = \cos(\frac{v}{2}), \qquad \phi_1 = \sin(\frac{v}{2}) \quad \text{and} \quad \phi_{-1} = -\sin(\frac{v}{2}), \qquad (6)$$

are light amplitudes in the zero-th, plus first and minus first diffraction order, respectively and asterisk * stands for complex conjugate. (When light impinges an ultrasonic wave at an angle which does not exactly fulfil the Bragg condition, Eq. (4), or for insufficiently big values of the Klein-Cook parameter $Q$, the above amplitudes, Eq. (6), become described by complex functions).

In case of $Q \approx 1$ one can speak about the so-called "transition region" or "the intermediate range" which covers light diffraction phenomena between the described above two limiting cases. As opposed to the Raman-Nath and Bragg diffraction, the phase of diffracted light becomes within the transition region a function of the Raman-Nath and Klein-Cook parameters [16, 17]. It should be emphasized that regardless of the diffraction regime, the angular frequency of light in the n-th diffraction order, $\omega_n$, is changed due to the Doppler effect, and it can be expressed by a simple formula: $\omega_n = \omega + n\Omega$, where $\omega$ is the angular frequency of incident light on ultrasonic wave and $\Omega$ denotes the angular frequency of ultrasound.

## 3. Interaction of Photon Pairs with Ultrasonic Wave

The quantum picture of light diffraction is usually formulated in single photons language. A light beam can be treated as a system of independent photons following the same probability distribution determined by a single photon wave function. The diffracting waves are solutions of the Maxwell equations (in our case the Raman-Nath system [12]) and can be interpreted as single photon wave functions [18]. In case of light beams consisting of photon pairs formulas describing diffraction phenomenon should be appropriately modified.

Considering the Bragg diffraction regime, one can utilize Eq. (6) to derive relevant probability amplitudes of finding photons in the zero-th and first diffraction order in case of collinear correlated photon pairs irradiating an ultrasonic wave at a Bragg angle. The probability amplitude that the first photon of the incidenting photon pair will, after the interaction with an ultrasonic wave, be detected in the first $\Psi_1^1$ or the zero-th $\Psi_0^1$ diffraction order, respectively, can be expressed as:

$$\Psi_1^1 = \sin(\frac{v}{2}), \qquad \Psi_0^1 = \cos(\frac{v}{2}). \qquad (7)$$

Analogically, the probability amplitude that the second photon of the incidenting pair will, after the interaction with an ultrasonic wave, be detected in the first $\Psi_1^2$ or the zero-th $\Psi_0^2$ diffraction order can be written as:

$$\Psi_1^2 = \sin(\frac{v}{2}), \qquad \Psi_0^2 = \cos(\frac{v}{2}). \qquad (8)$$

Consequently, the probability amplitude that both the first and the second photon of a considered photon pair will be found in the first $\Psi_1^{1,2}$ or the zero-th $\Psi_0^{1,2}$ diffraction order equals the product of appropriate probability amplitudes:

$$\Psi_1^{1,2} = \sin^2(\frac{v}{2}), \qquad \Psi_0^{1,2} = \cos^2(\frac{v}{2}). \qquad (9)$$

Accordingly, the probability of simultaneous detection of two photons in the first or the zero-th diffraction order, respectively, can be expressed as:

$$P_1 = \sin^4(\frac{v}{2}), \qquad P_0 = \cos^4(\frac{v}{2}). \qquad (10)$$

It follows from the last equation that in case of a light beam consisting of collinear correlated photon pairs, the resulting light intensity distribution in the zero-th diffraction order will vary as $I_0 = \cos^4(\frac{v}{2})$ and in the first diffraction order as $I_1 = \sin^4(\frac{v}{2})$. Thus, while comparing this situation to ultrasonic diffraction of single photons, Eq. (5), one can see that the diffracted light intensity maxima and minima appear, in both cases, for the same values of the Raman-Nath parameter, and only the shape of relevant functions is different.

While making use of Eqs. (7) and (8) we can now calculate the probability amplitude $\Psi_{0,1}^{1,2}$ of two simultaneous detections of a single photon in both the zero-th and the first diffraction orders:

$$\Psi_{0,1}^{1,2} = \Psi_0^1 \cdot \Psi_1^2 = \Psi_0^2 \cdot \Psi_1^1 = \frac{1}{2}\sin(v). \qquad (11)$$

Therefore, the probability $P_{0,1}$ of two simultaneous photon detections in both the zero-th and the first diffraction orders can be written as:

$$P_{0,1} = \frac{1}{4}\sin^2(v). \qquad (12)$$

From the Eqs. (11) and (12) it is clear that the probability $P_{0,1}$ of two simultaneous photon detections in the zero-th and first diffraction orders is also a periodical function of the Raman-Nath parameter value, which period is twice shorter than the one describing single photon counts in separate diffraction orders, Eq. (5).

**4. Experimental Arrangement**

Fig. 2 shows a scheme of the experimental set-up employed for investigation of the diffraction phenomenon of collinear correlated photon pairs on an ultrasonic wave within the Bragg regime. This arrangement is similar to the one which was utilized for examination of ultrasound diffraction of collinear correlated photon pairs within the Raman-Nath and intermediate regime [8].

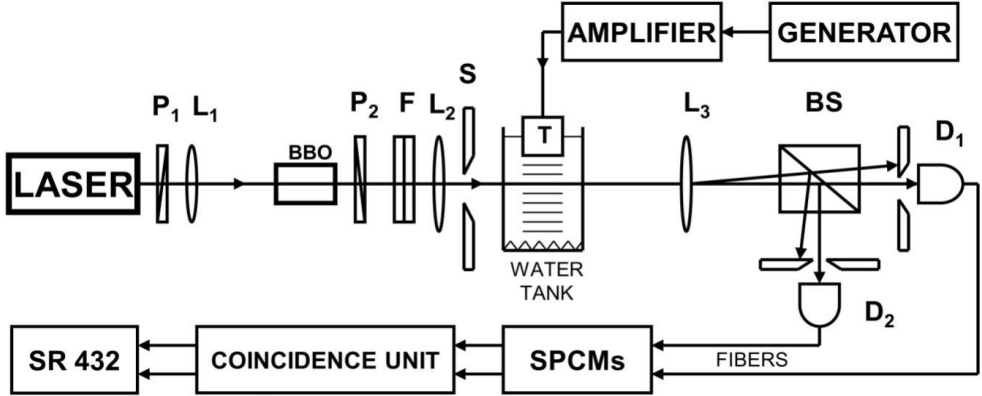

FIG. 2. Scheme of the experimental setup [8]. Major components include an ultraviolet laser, down-conversion crystal (BBO), polarizers ($P_1$, $P_2$), lenses ($L_1 - L_3$), optical filters set (F) (long-pass and band-pass filter), slit (S), water tank with an ultrasonic transducer (T), nonpolarizing beam splitter (BS), two sets for the selection of diffraction orders, equipped with relevant optics directing photons into optical fibers ($D_1$, $D_2$), single-photon counting modules (SPCMs) and dual channel photon counter (SR 432). Focal lengths of the $L_1$, $L_2$ and $L_3$ lenses equalled 500, 158 and 364 millimetres, respectively.

Focussed light from an InGaN laser diode (405 nm) is sent through a BBO crystal producing photon pairs via type-I parametric down-conversion. The crystal was oriented for collinear down-conversion of the pairs with equal wavelengths (810 nm). The crystal set between two crossed polarizers $P_1$ and $P_2$. The first polarizer $P_1$ ensures that pump photons are linearly polarized and retain an appropriate polarization direction with respect to the optical axis of the BBO crystal. The second polarizer $P_2$ eliminates the pump beam, which has polarization orthogonal to that of down-converted photons. For better separation from the pump laser photons, a supplementary long-pass filter (RG780) is introduced, which passes wavelengths longer than 780 nm. While the aim of the band-pass filter is to pass further only the photons of 810 nm wavelength with 10 nm bandwidth (FWHM). A set of $L_1$ and $L_2$ lenses ensures a precisely parallel beam of the correlated photon pairs. In fact, besides the photon pairs there are also single photons in the beam. To obtain photon pairs of a given wavelength and propagating in a specific direction, relevant apertures and optical filters have to be introduced behind the nonlinear crystal. In the process of spatial and temporal filtration of photon pairs, numerous photons from individual photon pairs are lost and, in such a way, single photons are generated too. Consequently, there are also single photons present in the produced beam and, what is more, the number of single photons is always considerably higher than that of correlated photon pairs [8,20]. The down-converted photons pairs and single photons incident, after passing through a rectangular slit S (2×1.5 mm), on a progressive ultrasonic wave propagating in a water-tank. The ultrasonic field was generated using a circular LiNbO$_3$ transducer

T with fundamental frequency of 10.790 MHz, 22.4 mm in diameter, with coaxial electrodes, excited at a frequency of 32.370 MHz which corresponds to the 3$^{rd}$ harmonic. Standing waves in the water-tank were avoided by careful selection of a suitable absorbing material. The angle between the propagation direction of the ultrasonic wave and the direction of the down-converted photons pairs and single photons was tuned by changing the inclination of the ultrasonic transducer. Thanks to such technical arrangement, the water-tank remained steady during the adjustment of the incidence angle of photons irradiating ultrasonic wave. And, as an entrance window of the water-tank was illuminated normally, there was no need to account for the light beam refraction introduced by the water-tank. Photons pairs and single photons, after passing through the ultrasonic wave and a lens $L_3$, are split by a non-polarizing 50%/50% beam splitter BS into two beams. While examining coincidences of photons in the beams emerging from the beam splitter BS, one can determine whether the ultrasonic wave interacted with a single photon or a pair of photons. Spatially separated diffraction orders appear in the focal plane of the $L_3$ lens, associated with each of the mentioned beams. Photons from the two beams, after selection of desired diffraction order, are directed to a single-photon counting modules SPCMs by multi-mode optical fibers, equipped with appropriately chosen input diaphragms and lenses $D_1$ and $D_2$. Electrical signals from the SPCMs, after passing through a coincidence unit with coincidence window of 10 ns, were registered by a dual channel photon counter SR 432. The SPCMs had dark count rates of about 450 counts per second (cps). The experiments were carried with the laser output power of $(25.0 \pm 0.1)$ mW which corresponded to approximately 100 000 cps in each of the photon beams. While the number of coincidences was about 2 100 cps. When the set-up registers coincidences in one selected diffraction order (or coincidences between two different diffraction orders) we know that a collinear photon pair interacted with the ultrasonic wave. Whereas the total number of counts related to only one of the photon beams, decreased by: the number of coincidences, the number of counts corresponding to photon pairs not split by a beam splitter and the number of dark counts, corresponds to the number of single photons interacting with the ultrasonic wave. Of course, one should consider here also the so-called accidental coincidences which increase the number of coincidences being registered. The number of accidental coincidences is calculated from the number of counts registered at the inputs of the coincidence unit and the width of the coincidence window.

## 5. Experimental Results

At the very beginning we examined whether the positive and negative Bragg angles for collinear correlated photon pairs diffraction equal the ones corresponding to single photons. For this reason, having set a fixed value of the Raman-Nath parameter, the angle of photons incidence on an ultrasonic wave was changed and the related numbers of single photon and photon pair counts in the zero-th diffraction order were recorded. The minima of the registered photon counts correspond to Bragg angles.

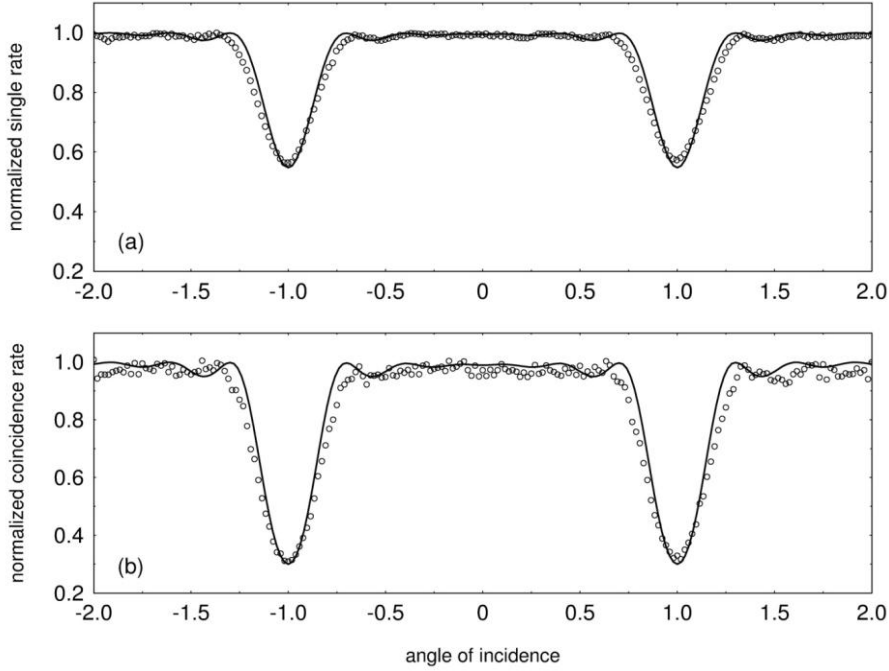

FIG. 3 Theoretical expectation (solid line) and experimental (circles) normalized counts as a function of the photons incidence angle on the ultrasonic wave, expressed in units of the Bragg angle, for the Raman-Nath parameter $v = 1.47$ and the Klein-Cook parameter $Q = 40.82$. Fig. 3a illustrates the case of single photons diffraction, while Fig. 3b corresponds to collinear correlated photon pairs diffraction.

In Fig. 3 we show the number of registered single photons (Fig. 3a) and photon pairs (Fig. 3b) in the zero-th diffraction order, plotted as a function of the photons incidence angle $\Theta$ on the ultrasonic wave, expressed in units of the Bragg angle. The data was normalized to unity by taking the number of single photon counts of 247 400 and photon pairs of 5 440 equal one, respectively, which corresponded to the situation of ultrasound being switched off. (The data was corrected for all the discussed earlier factors, including accidental coincidences). The angle $\Theta$ was changed within the range from $-2 \times \Theta_B$ to $+2 \times \Theta_B$, where $\Theta_B = \arcsin(\frac{\lambda}{2\mu_0 \Lambda})$ defines the Bragg angle for single photons (Eq. (4)).

As can be seen from Figs. 3a and 3b, minima of the zero-th order diffracted light intensities show for exactly the same values of photons incidence angle for both single photons and photon pairs. Appropriate profile of the zero-th diffraction order light intensity distribution for single photons was derived numerically from the relevant Raman-Nath equations set [12] by means of the NOA method ($N^{th}$ Order Approximation Method) [16, 19], while taking for calculations diffraction orders from $-4$ up to $+4$ (i.e., $N=4$). Whereas the corresponding profile for photon pairs was obtained by taking the square of light intensity distribution describing single photons diffraction, according to the idea discussed while deriving Eq. (10). It should be clearly emphasized that the Bragg angles for single photons and photon pairs diffraction are identical. Having proved that the Bragg angle is the same for both photon pairs and single photons diffraction, we set about examining the number of single photons (Eq. (5)) and photon pairs (Eq.(10)) in the zero-th diffraction order as a function of the Raman-Nath parameter.

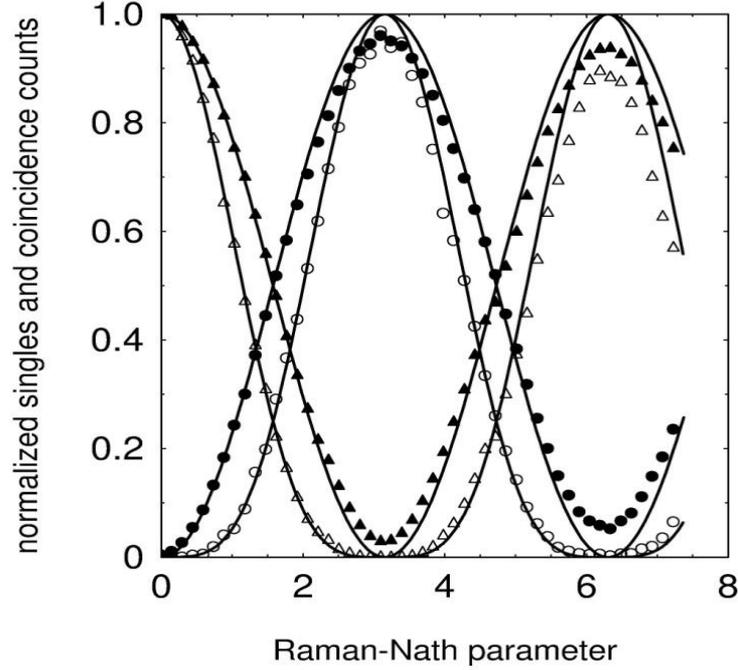

FIG. 4. Experimental data corresponding to diffracted light intensity distribution in the zero-th (triangles) and first (circles) diffraction orders for Bragg incidence, plotted as a function of the Raman-Nath parameter for single photons (filled symbols) and collinear correlated photon pairs (hollow symbols). Relevant theoretical curves for single photons, Eq. (5), and for collinear photon pairs, Eq. (10), are represented by solid lines.

In Fig. 4 a comparison is presented of theoretical predictions based on Eqs. (5) and (10) with relevant experimental data for the Bragg diffraction regime, for the zero-th and first diffraction orders. The Raman-Nath parameter was varied from zero up to a value slightly above $2\pi$, so as to cover a complete period of diffracted light intensity changes. Also in this case, the experimental data for both single and coincidence counts was normalized to unity by taking as one the number of photon counts registered with the ultrasonic wave switched off.

The number of registered counts corresponding to the Raman-Nath parameter value equal zero (ultrasound switched off) was 439 316 for single and 8 980 for coincidence counts, respectively. These values were used to normalize data in the zero-th and next in the first diffraction order. (Of course, the data was again corrected for all the discussed earlier factors, including accidental coincidences).

In view of the results depicted in Fig. 4 it can be seen that the period of photon counts variation with the Raman-Nath parameter changes is the same for single photons and for photon pairs diffraction by an ultrasonic wave. Moreover, Fig. 4 reveals that as the Raman-Nath parameter value raises, increases the discrepancy between experimental data and relevant theoretical curves based on Eqs. (5) and (10). Indeed, already for the Raman-Nath parameter $v = \pi$ the number of photons neither drops to zero in the zero-th diffraction order nor it reaches one in the first diffraction order. For $v = \pi$ even bigger discrepancy is evident. The explanation of the observed tendency is a too small number of the ultrasonic wave wavelengths contributing to the registered image of diffracted light intensity distribution. The mentioned number is limited by a size of the opening in the diaphragm placed in front of the water tank, which in our case had 2 mm along the direction of ultrasound propagation, which covered 44 wavelengths. To observe experimentally a complete fading of light in the zero-th diffraction order or to reach , at the same time, light intensity equal unity within the first diffraction order for Raman-Nath parameter $v = \pi$, one would need involvement of considerably higher number of the ultrasonic wave wavelengths than it was assured in our experiment. Of course, it is possible to increase the diaphragm opening size (along the direction of ultrasound propagation) so as to include bigger number of the ultrasonic wave wavelengths but then one should take into account also the attenuation coefficient of the ultrasound, that is, the Raman-Nath parameter value changes along the direction of ultrasonic wave propagation. Therefore, a compromise between the number of ultrasonic wave wavelengths and ultrasound attenuation was decided while selecting the diaphragm opening size.

In the next figure, Fig. 5, the experimental results are shown illustrating the registered number of coincidences between photons from the zero-th and first diffraction orders and, additionally, the number of single photons from the zero-th diffraction order. While making use of the number of single photons detected in the zero-th diffraction order ( 886 904 counts for $v = 0$ ) and considering the determined experimentally efficiency of our source of photon pairs (2.04±0.02)%, one could normalize the experimental data of the registered coincidences between photons from the zero-th and first diffraction order.

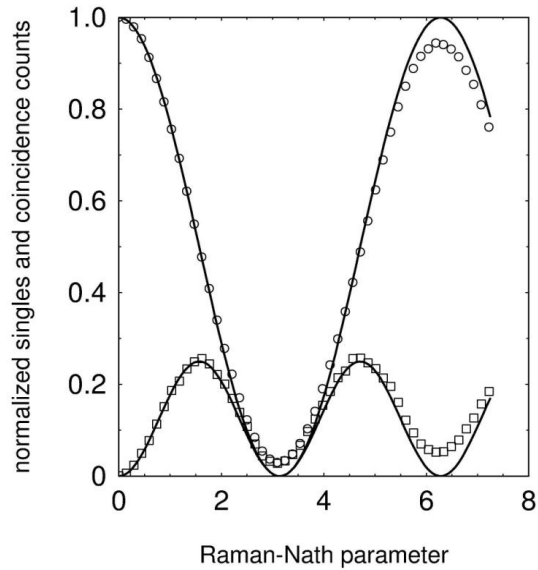

FIG. 5 Theoretical expectation, according to Eq. (12), (solid line) and experimental data (squares) for normalized coincidence counts between photons in the zero-th and first diffraction order, shown as a function of the Raman-Nath parameter. For comparison, the circles and the other solid line show experimental data and relevant theoretical curve, Eq. (5), respectively, obtained for single photon counts in the zero-th diffraction order.

The experimental data corresponding to the number of coincidences between photons from the zero-th and first diffraction orders are compared to relevant theoretical curve based on Eq. (12). A good agreement is visible. From Fig. 5 and Eq. (12) one can easily see that the period of changes of the registered photon coincidences between the zero-th and first diffraction order, examined as a function of the Raman-Nath parameter value, is twice shorter than the one corresponding to the number of single photons detected in the zero-th diffraction order.

## 6. Conclusions

A simple model of the interaction of collinear correlated photon pairs with ultrasonic waves, presented in the present paper reveals very good agreement with corresponding experimental data. This model has already been successfully applied also in case of experimental investigation of ultrasound diffraction of collinear correlated photon pairs within the Raman-Nath and intermediate light diffraction region [8], showing good agreement between experimental data and relevant theoretical predictions. In view of the obtained results it should be possible to calculate required or desirable parameters of acousto-optic devices illuminated with collinear photon pairs or beams consisting of both pairs of photons and single photons. These results are promising and open the doors to investigation of photon pairs diffraction by an ultrasonic wave in media with ultrasonically induced optical birefringence [21], which may potentially enable fast manipulation of polarization state of photon pairs. Moreover, it is worth noting that the discussed in our paper description of collinear correlated photon pairs interaction with ultrasonic waves should be very useful also in case of examination of the diffraction phenomenon of photon pairs by adjacent [22, 23 ] and superposed [24 ] ultrasonic beams.

**Acknowledgement**

I thank Professor M. Żukowski for his advice and inspiration. This work was supported by the National Science Centre research grant no N 0855/B/H03/2011/40.